\documentclass[useAMS,usenatbib]{mn2e}
\usepackage{graphicx,natbib,times,amssymb}
\title[The effects of SPS on the distributions of
the asteroseismic observables $\nu_{max}$ and $\Delta\nu$] {The
effects of stellar population synthesis on the distributions of the
asteroseismic observables $\nu_{max}$ and $\Delta\nu$ of red-clump
stars}
\author[W. Yang, X. Meng and Z. Li]{Wuming Yang$^{1,2}$\thanks{E-mail:
yangwuming@ynao.ac.cn}, Xiangcun Meng$^{1}$ and Zhongmu Li$^{3,4}$ \\
$^{1}$School of Physics and Chemistry, Henan Polytechnic University,
Jiaozuo 454000, Henan, China.\\
$^{2}$Department of Astronomy, Beijing Normal University, Beijing 100875, China.\\
$^{3}$Institute for Astronomy and History of Science and Technology,
Dali University, Dali 671003, China.\\
$^{4}$National Astronomical Observatories, Chinese Academy of
Sciences, Beijing 100012, China.}
\begin{document}

\date{ }

\pagerange{\pageref{firstpage}--\pageref{lastpage}} \pubyear{2010}

\maketitle

\label{firstpage}

\begin{abstract}
The distributions of \textbf{the} frequencies of the maximum
oscillation power ($\nu_{max}$) and the large frequency separation
($\Delta\nu$) of red giant stars observed by CoRoT have a dominant
peak. Miglio et al. \textbf{(2009) identified} that the stars are
red-clump stars. Using stellar population synthesis method, we
studied the effects of Reimers mass loss, binary interactions, star
formation rate and the mixing-length parameter on the distributions
of \textbf{the} $\nu_{max}$ and $\Delta\nu$ of red-clump stars. The
Reimers mass loss can result in an increase in the $\nu_{max}$ and
$\Delta\nu$ of old population which lost a considerable amount of
mass. \textbf{However, it leads} to a small decrease in those of
middle-age population which lost a little bit of mass.
\textbf{Furthermore, a} high mass-loss rate impedes the low-mass and
low-metal stars evolving into core-helium burning \textbf{(CHeB)}
stage. Both Reimers mass loss and star formation rate mainly affect
the number of CHeB stars with $\nu_{max}$ and $\Delta\nu$, but
\textbf{hardly affect} the peak locations of $\nu_{max}$ and
$\Delta\nu$. Binary interactions also can lead to\textbf{ an
increase or decrease in the $\nu_{max}$ and $\Delta\nu$ of some
stars}. \textbf{However,} the fraction of CHeB stars undergoing
binary interactions is very small in our simulations.
\textbf{Therefore,} the peak locations are also not affected by
binary interactions. The non-uniform distributions of $\nu_{max}$
and $\Delta\nu$ are mainly caused by the most of red-clump stars
having an approximate radius rather than mass. \textbf{The} radius
of red-clump stars decreases with increasing the mixing-length
parameter. \textbf{The} peak locations of $\nu_{max}$ and
$\Delta\nu$ can, \textbf{thus}, be affected by the mixing-length
parameter.

\end{abstract}


\section{Introduction}
Solar-like oscillations have been confirmed in many main-sequence
and subgiant stars, such as $\alpha$ Cen A \citep{bedd04}, $\alpha$
Cen B \citep{kjel05}, Procyon A \citep{egge04}, $\eta$ Bootis
\citep{carr05}, etc. Some giant stars, such as $\xi$ Hya
\citep{fran02}, $\eta$ Ser \citep{barb04}, and $\epsilon$ Oph
\citep{deri06}, have also been \textbf{found the behaviour of
oscillating}. Asteroseismology is a powerful tool for determining
the fundamental parameters of individual stars \citep{egge05,
egge06, yang10}, and it has significantly advanced the
\textbf{theories of stellar structure and stellar evolution}.

\begin{figure}
     \includegraphics[angle=-90, width=8cm]{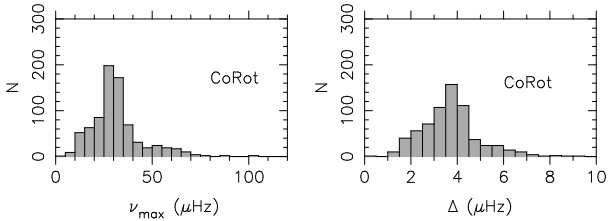}
     \centering
       \caption{The histogram of the $\nu_{max}$ and $\Delta\nu$
        obtained by \citet{hekk09}.}
       \label{fcor}
   \end{figure}

The detection of solar-like oscillations in red giant stars opened
up a new field that can be explored with asteroseismic techniques.
For an ensemble of cluster stars all having the same age and
metallicity, studying oscillations in the cluster stars should
provide more constraints on \textbf{the theory of stellar evolution}
than just fitting parameters of single oscillating stars
\citep{stel07}. Thus a number of attempts have been made to detect
solar-like oscillations in giant stars of clusters \citep{gill93,
gill08, edmo96, stel07, fran07, stel09, stel10, bedd10}. The
oscillations of giant stars have been detected in a few clusters,
such as open cluster M67 \citep{stel07}, globular cluster 47 Tucanae
\citep{edmo96} and NGC 6397 \citep{stel09}. However, these
observations did not obtain oscillation frequencies. Recently, using
the data observed by the first CoRoT \citep{bagl06} 150-day long run
in the direction of the galactic centre (LRc01), \citet{hekk09}
obtained the frequencies of maximum oscillation power ($\nu_{max}$)
and the large separations ($\Delta\nu$) of about 800 solar-like
oscillating red giants. They found that distributions of the
$\nu_{max}$ and $\Delta\nu$ are non-uniform (see Fig. \ref{fcor}),
which provide an opportunity for detailed studies of population of
galactic-disk red giants \citep{migl09}. Additionally, theoretical
$\nu_{max}$ and $\Delta\nu$ can be calculated using equations
\citep{kjel95}
   \begin{equation}
    \nu_{max}=3050 \frac{M/M_{\odot}}{(R/R_{\odot})^{2}\sqrt{T_{eff}/5777 K}}\mathrm{\mu Hz}\,,
    \label{eq1}
   \end{equation}
   and
   \begin{equation}
    \Delta \nu=134.9 \frac{(M/M_{\odot})^{1/2}}{(R/R_{\odot})^{3/2}} \mathrm{\mu
    Hz}\,. \label{eq2}
   \end{equation}
Thus, we can obtain the theoretical distributions of $\nu_{max}$ and
$\Delta\nu$ by simulating the distributions of stellar mass, radius
and effective temperature.

Using the stellar population synthesis code TRILEGAL, \citet{migl09}
identified that the oscillating giants observed by CoRoT are
primarily red-clump stars, i.e. post-flash core-He-burning (CHeB)
stars, and found that theoretical distributions of $\nu_{max}$ and
$\Delta\nu$ are good global agreement with observed ones. Equations
\ref{eq1} and \ref{eq2} show that the theoretical distributions of
$\nu_{max}$ and $\Delta\nu$ are relative to the mass and radius
distributions of population. \textbf{Therefore,} \cite{migl09}
suggested that mass-loss rate during the red-giant branch (RGB) and
star formation rate may affect the distributions of the $\nu_{max}$
and $\Delta\nu$.

In this paper, we focused the effects of mass loss, binary
interactions, the mixing-length parameter, etc. on \textbf{the
distributions of $\nu_{max}$ and $\Delta\nu$ of CHeB stars}. We used
the Hurley rapid single and binary evolution codes \citep{hurl00,
hurl02} to construct stellar models. In Hurley's codes, mass-loss
efficiency is an adjustable parameter. They can, \textbf{thus}, be
applied to studying the effect of mass loss on \textbf{the
distributions}.

The paper is organized as \textbf{follows}. We simply show our
stellar population synthesis method in section 2. The results are
represented in section 3. In section 4, we discuss and summarize the
results.

\section{Stellar population synthesis}

To simulate the stellar population observed $\nu_{max}$ and
$\Delta\nu$ \citep{hekk09}, we calculated single-star stellar
population (SSP) and binary-star stellar population (BSP),
respectively. Stellar samples are generated by the Monte Carlo
simulation. The basic assumptions for the simulations are as
follows. (i) Star formation rate (SFR) is assumed to be a constant
over the past 15 Gyr. (ii) The age-metallicity relation is taken
from \cite{roch00a}. (iii) The lognormal initial mass function (IMF)
of \cite{chab01} is adopted. Additionally, for BSP, we generate the
mass of the primary, $M_{1}$, according to the IMF. The ratio ($q$)
of the mass of the secondary to that of the primary is assumed to be
an uniform distribution within 0-1 for simplicity. Then the mass of
the secondary star is given by $qM_{1}$. We assume that all stars
are members of binary systems and that the distribution of
separations is constant in $\log a$ for wide binaries and falls off
smoothly at close separation:
\begin{equation}
an(a)=\left\{
 \begin{array}{lc}
 \alpha_{\rm sep}(a/a_{\rm 0})^{\rm m} & a\leq a_{\rm 0};\\
\alpha_{\rm sep}, & a_{\rm 0}<a<a_{\rm 1},\\
\end{array}\right.
\end{equation}
where $\alpha_{\rm sep}\approx0.070$, $a_{\rm 0}=10R_{\odot}$,
$a_{\rm 1}=5.75\times 10^{\rm 6}R_{\odot}=0.13{\rm pc}$ and
$m\approx1.2$. This distribution implies that the \textbf{number} of
wide binary system per logarithmic interval \textbf{is} equal, and
that approximately 50\% of the stellar systems are binary systems
with orbital periods less than 100 yr \citep{HAN95}. \textbf{With
these assumptions}, we calculated the evolutions of $10^{6}$ stars
with an initial mass between 0.8 and 5.8 $M_{\odot}$ and
mixing-length parameter $\alpha$ = 2.0. The theoretical $\nu_{max}$
and $\Delta\nu$ of single and binary stars are calculated using
equations (\ref{eq1}) and (\ref{eq2}).

\section{Calculation results}

\subsection{Mass-loss effect}
\begin{figure}
\begin{center}
\includegraphics[angle=-90, width=0.47\textwidth]{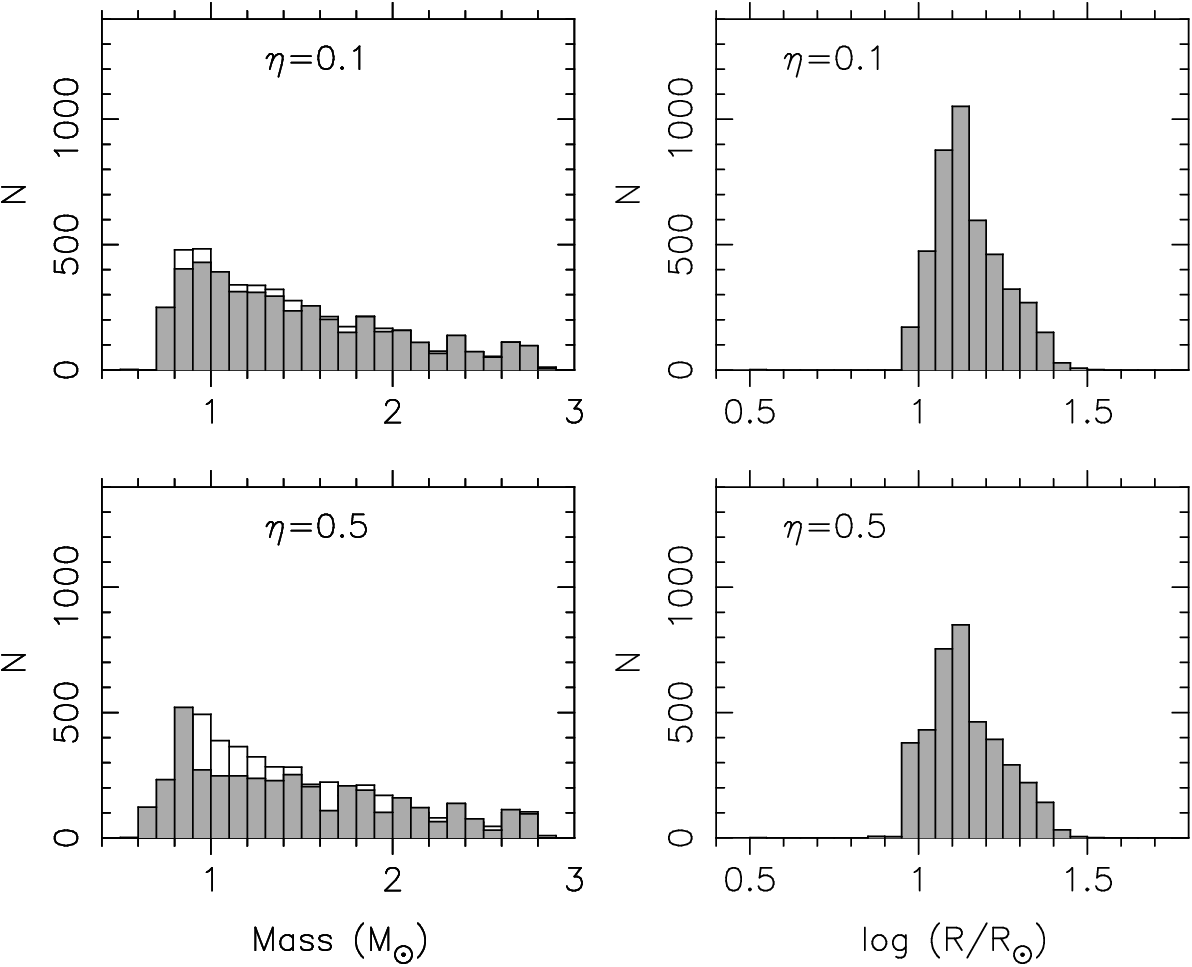}
\includegraphics[angle=-90, width=0.47\textwidth]{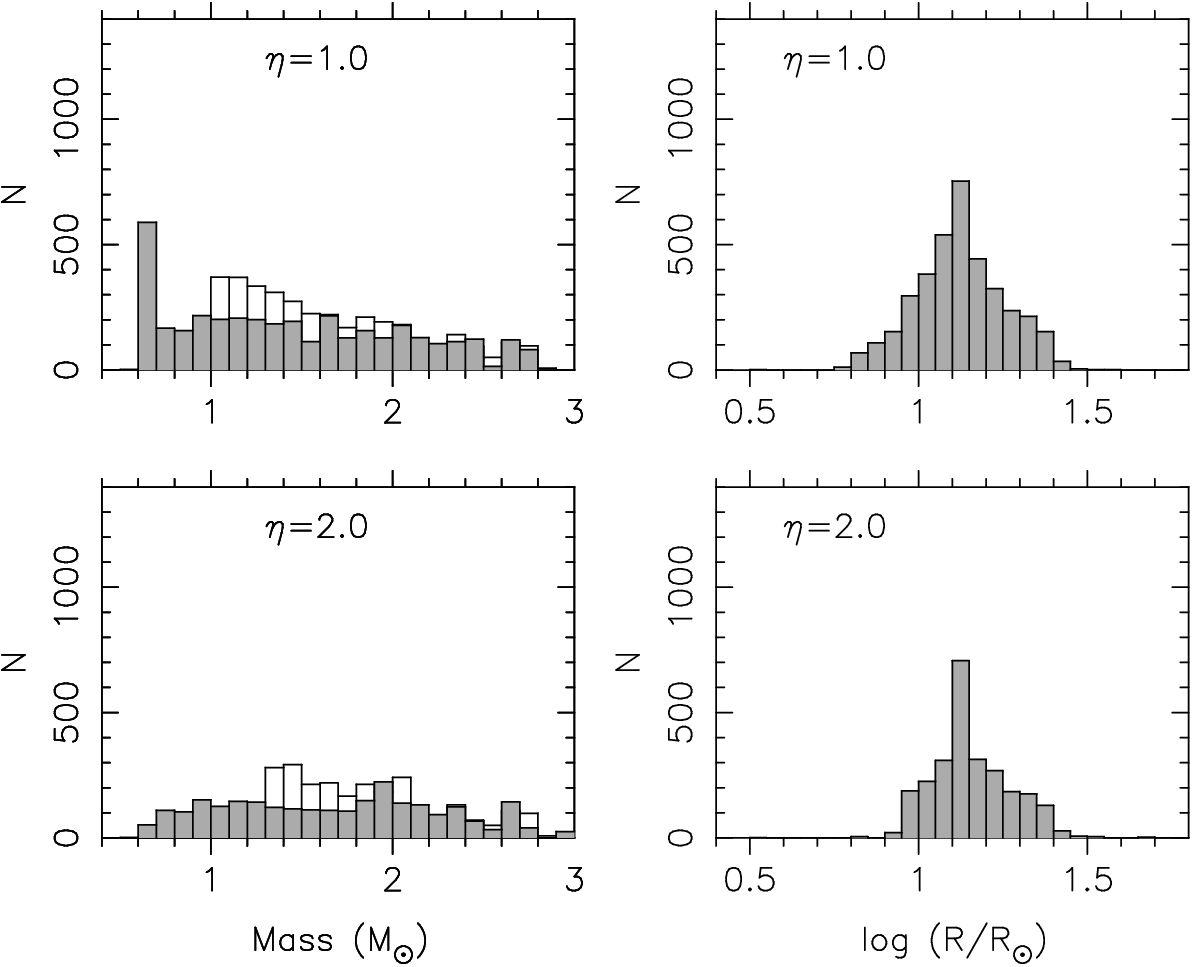}
 \caption{The histograms of the mass and radius of the CHeB stars of
 simulated SSP with different mass-loss coefficient $\eta$.
 In the left panels, empty bars illustrate the mass distribution of
 the progenitors of the CHeB stars.}
 \label{fmrd}
\end{center}
\end{figure}

Fig. \ref{fmrd} shows the distributions of \textbf{the} mass and
radius of \textbf{the} CHeB stars of simulated SSP with different
Reimers \citep{reim75} mass-loss efficiency ($\eta$). The
distributions of \textbf{the} initial mass of \textbf{the}
progenitors of the CHeB stars also are shown in Fig. \ref{fmrd}.
Reimers mass loss mainly affects the stars with initial mass less
than 2 $M_{\odot}$. With the increase in the $\eta$, the mass lost
by a star during the RGB increases, which leads to many low-mass
stars can not evolve into the CHeB stage, and the lower limit of
\textbf{the} initial mass of stars being able to evolve into the
CHeB stage increases too. For example, when the value of $\eta$
increases from 1.0 to 2.0, the lower limit increases from about 0.95
$M_{\odot}$ to 1.25 $M_{\odot}$. Thus the fraction of low-mass CHeB
stars and the sample size of the simulated CHeB stars decreases with
increasing $\eta$. For $\eta$ = 2.0, the mass distribution is almost
uniform. Moreover, the distribution of logarithmic radius has a
fixed peak location between 1.05 and 1.15, which is almost not
affected by the mass loss.

\begin{figure}
\begin{center}
\includegraphics[angle=-90, width=0.47\textwidth]{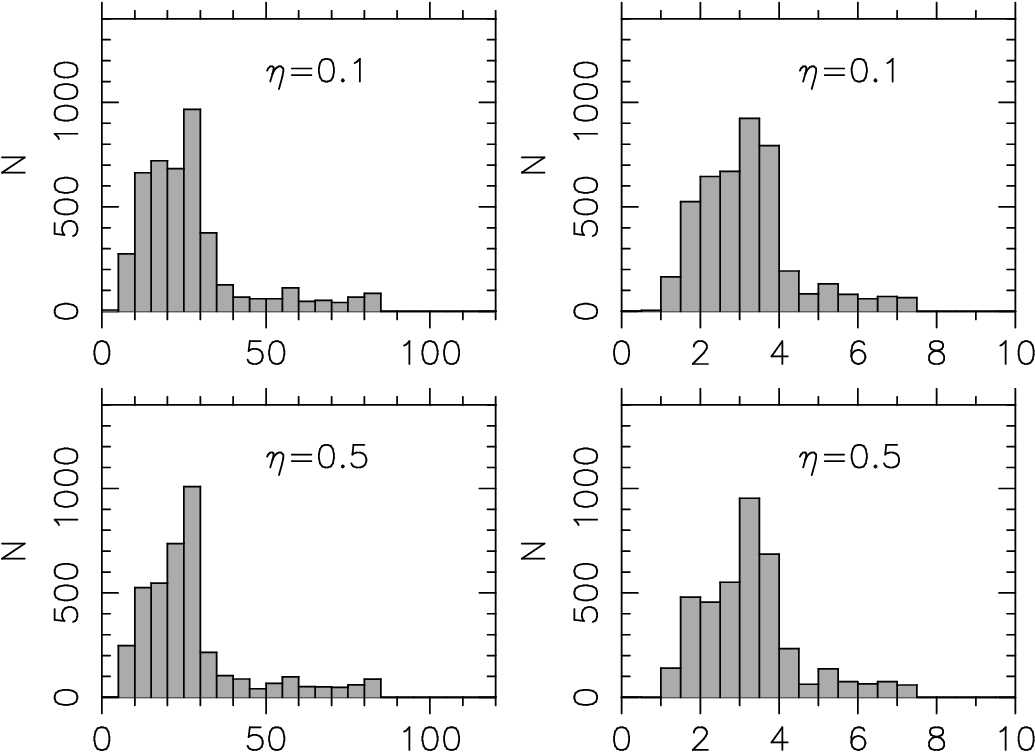}
\includegraphics[angle=-90, width=0.47\textwidth]{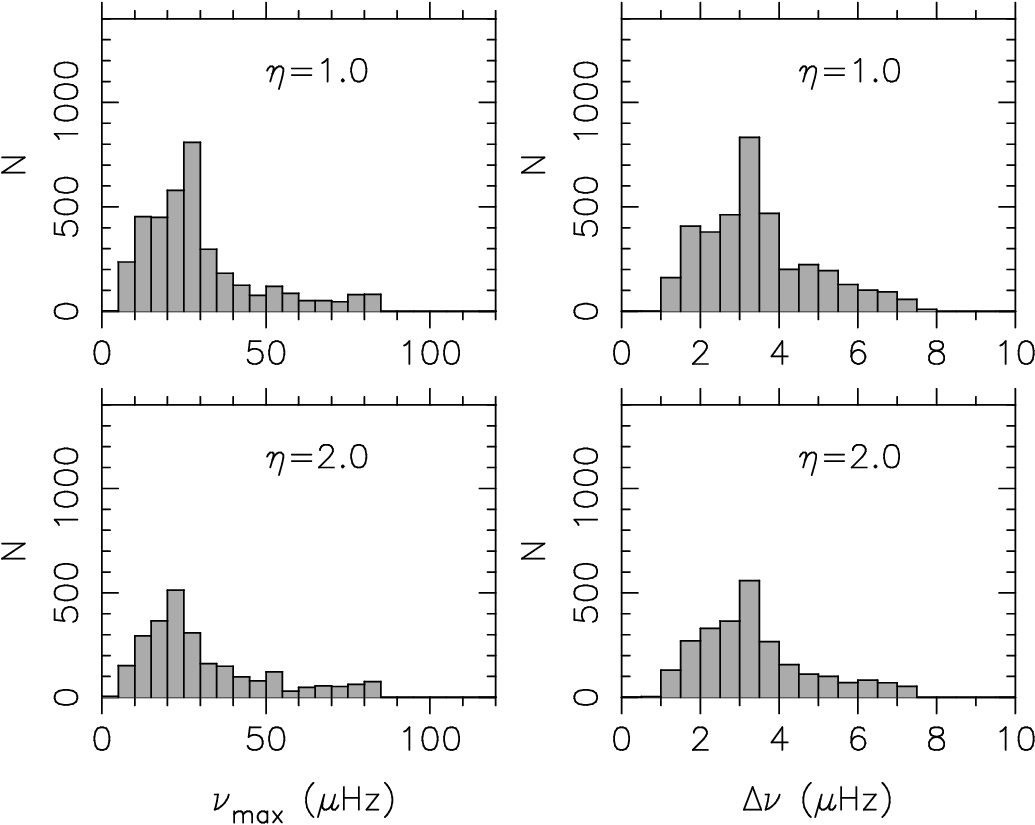}
 \caption{The histograms of \textbf{the} $\nu_{max}$ and $\Delta\nu$
 of \textbf{the} CHeB stars of simulated SSP with different
 mass-loss coefficient $\eta$.}
 \label{fmud}
\end{center}
\end{figure}

The distributions of \textbf{the} $\nu_{max}$ and $\Delta\nu$ of the
simulated CHeB stars are represented in Fig. \ref{fmud}. Both
distributions have a dominant peak. But the peak locations are
slightly lower than \textbf{the observed ones}. With the increase in
$\eta$, the number of the CHeB stars with $\nu_{max}$ \textbf{in the
range of about 10$\thicksim$30 $\mu$Hz }and $\Delta\nu$ \textbf{in
the range of about 1$\thicksim$4 $\mu$Hz} decreases. However, the
dominant peak for $\Delta\nu$ \textbf{always} locates between
\textbf{3 $\mu$Hz and 3.5 $\mu$Hz, which implies that} the effect of
the mass loss on the $\Delta\nu$ is not enough to change the peak
location. For $\eta \leq$ 1, the dominant peak of $\nu_{max}$
locates \textbf{between 25 $\mu$Hz and 30 $\mu$Hz}. However, when
the $\eta$ increases from 1.0 to 2.0, the peak location of
$\nu_{max}$ moves to \textbf{between 20 $\mu$Hz and 25 $\mu$Hz}.

\begin{figure}
\begin{center}
\includegraphics[angle=-90, width=0.47\textwidth]{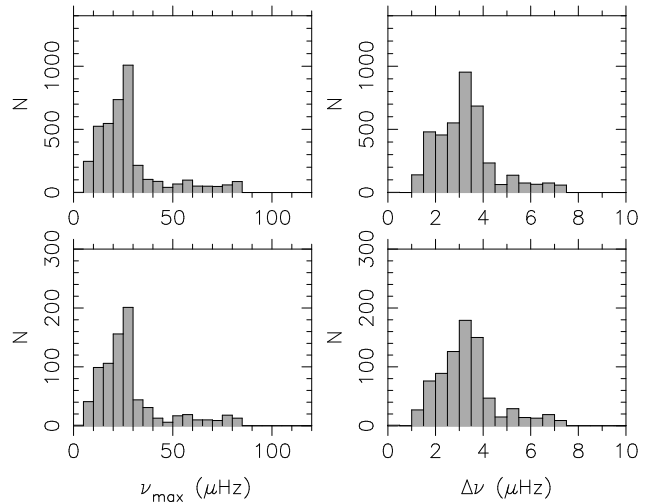}
 \caption{Histogram showing the effect of sample size on the
 distributions of \textbf{the} $\nu_{max}$ and $\Delta\nu$ of simulated CHeB stars.
 Upper panels show the results of 10$^6$ model evolutions. While lower panels
 represent the results of 2$\times10^5$ model evolutions. Both samples have
 the same other parameters. A K-S test shows that the discrepancies between
 two distributions are insignificant.}
 \label{fmtcl}
\end{center}
\end{figure}

The changes in the distributions of $\nu_{max}$ and $\Delta\nu$
\textbf{shown} in Fig. \ref{fmud} should not be due to the typical
variations in Monte Carlo simulations. In order to show the effect
of sample size on the distributions of $\nu_{max}$ and $\Delta\nu$,
we calculated the evolutions of $10^{6}$ stars which produced 3973
CHeB stars and the evolutions of 2$\times10^{5}$ stars which given
793 CHeB stars. Both samples have the same evolutionary parameters.
Figure \ref{fmtcl} shows that the distributions of $\nu_{max}$ and
$\Delta\nu$ of the two samples are very similar. A
Kolmogorov-Smirnov test shows that the discrepancies between the two
sample distributions are not significant: the Kolmogorov-Smirnov
statistic $D \simeq$ 0.030 for $\nu_{max}$ and 0.033 for
$\Delta\nu$, and the significance level of the $D$ is 0.61 for
$\nu_{max}$ and 0.45 for $\Delta\nu$.

\begin{figure}
\begin{center}
\includegraphics[angle=-90, width=0.47\textwidth]{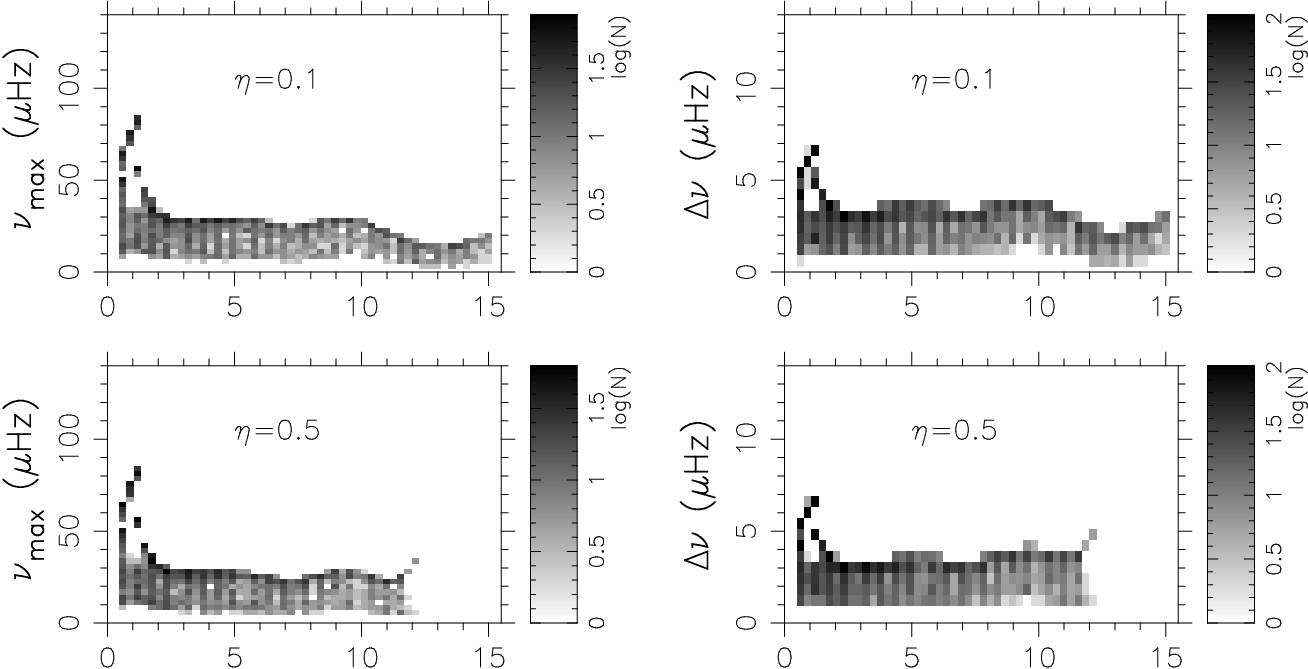}
\includegraphics[angle=-90, width=0.47\textwidth]{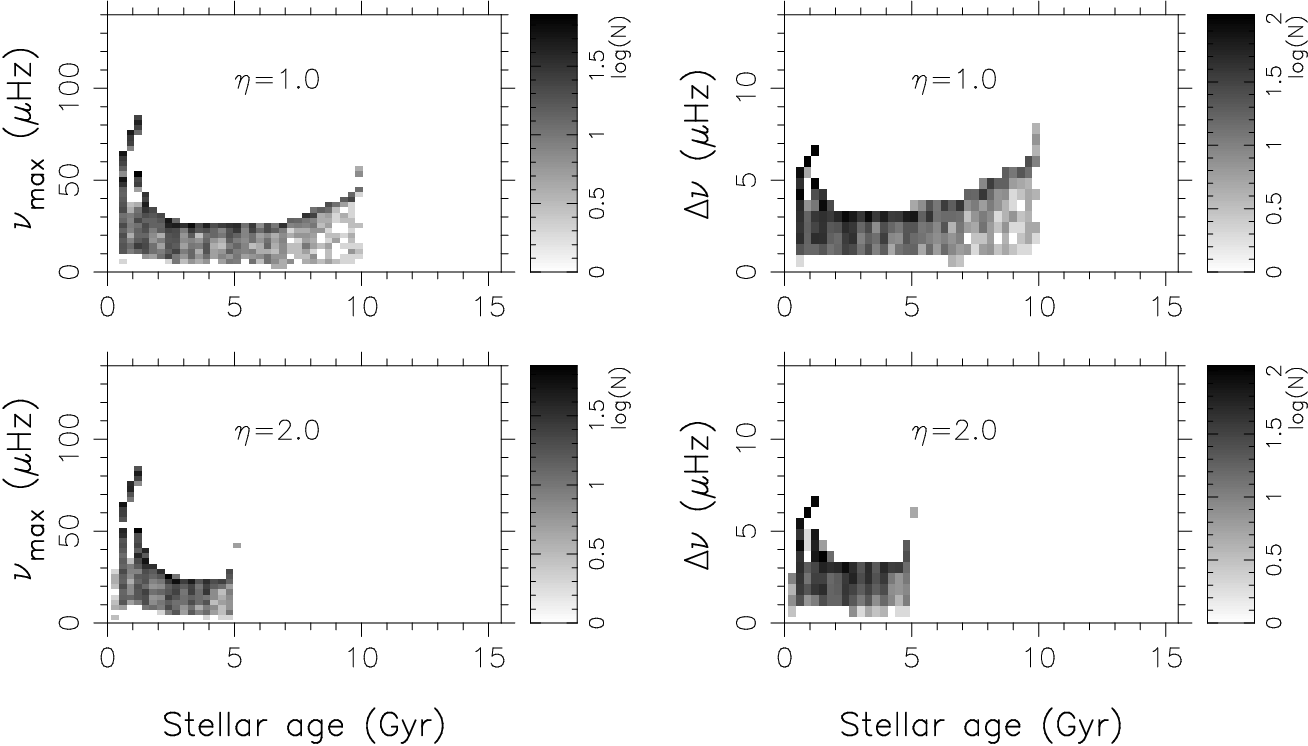}
 \caption{The distributions of \textbf{the} $\nu_{max}$ and $\Delta\nu$ of
 simulated CHeB stars with different $\eta$ as a function of age.}
 \label{faged}
\end{center}
\end{figure}

In Fig. \ref{faged}, we plotted the distributions of \textbf{the}
$\nu_{max}$ and $\Delta\nu$ of the simulated CHeB stars with
different $\eta$ as a function of stellar age. For age $>$ 2 Gyr,
Fig. \ref{faged} shows that the value of \textbf{the} $\nu_{max}$
and $\Delta\nu$ of CHeB stars is mainly located in the range of
$\sim$10 - $\sim$30 $\mu$Hz and $\sim$1 - $\sim$4 $\mu$Hz,
respectively. Moreover, most stars are located around the upper
boundary of the distributions. These stars are close to the zero-age
horizontal branch (ZAHB) in the Hertzsprung-Russell (HR) diagram.
While the stars around the lower boundary are approaching the
asymptotic giant branch (AGB). The CHeB stars with higher
$\nu_{max}$ and $\Delta\nu$ are mainly from the young population
with age less than 2 Gyr. Additionally, Fig. \ref{faged} clearly
indicates that Reimers mass loss mainly affects the $\nu_{max}$ and
$\Delta\nu$ of old stars. A big mass-loss rate impedes the low-mass
low-metal stars evolving into CHeB stage. \textbf{Consequently,} for
$\eta$ = 0.5, 1.0 and 2.0, the fraction of the CHeB stars with age
larger than 12, 10 and 5 Gyr is negligible, respectively. However,
the mass loss \textbf{scarcely affect} the distributions of
\textbf{the} $\nu_{max}$ and $\Delta\nu$ of stars with age $<$ 2
Gyr. When $\eta$ increases from 0.5 to 1.0, the upper boundary of
\textbf{the} $\nu_{max}$ and $\Delta\nu$ of the CHeB stars
\textbf{in the age range of about} 7-10 Gyr is enhanced, but the
effect of the increase in the $\eta$ on the distributions of
\textbf{the} $\nu_{max}$ and $\Delta\nu$ of the stars \textbf{in the
age range of} around 2-7 Gyr \textbf{is} not significant. When
$\eta$ increases to 2.0, the upper boundary of \textbf{the}
$\nu_{max}$ and $\Delta\nu$ of the stars with age approximately 5
Gyr \textbf{increases} too, \textbf{however}, the distributions of
\textbf{the} $\nu_{max}$ and $\Delta\nu$ of the stars \textbf{in the
age range of} about 2-4.5 Gyr move down slightly, which leads to the
peak location \textbf{in the $\nu_{max}$ histogram moving to 20-25
$\mu$Hz}.

\subsection{BSP effect}

\begin{figure}
\begin{center}
\includegraphics[angle=-90, width=0.47\textwidth]{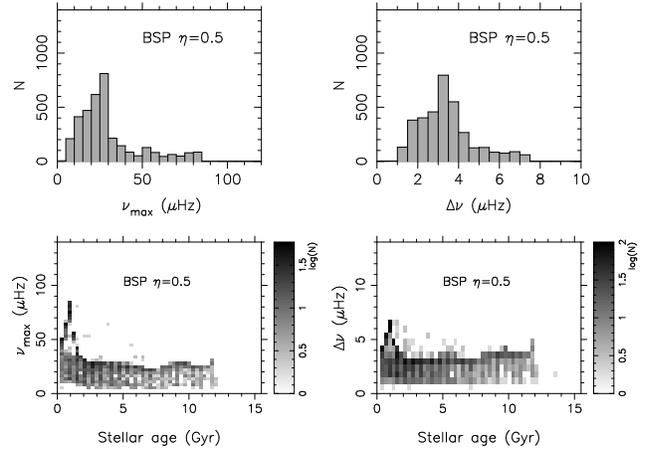}
 \caption{Upper panels show the histograms of \textbf{the} $\nu_{max}$ and
 $\Delta\nu$ of the CHeB stars of simulated BSP. Here, $\eta$ = 0.5.
 Lower panels indicate the distributions of the $\nu_{max}$
 and $\Delta\nu$ as a function of stellar age.}
 \label{fbsp}
\end{center}
\end{figure}

Fig. \ref{fbsp} shows the histograms of \textbf{the} $\nu_{max}$ and
$\Delta\nu$ of \textbf{the} CHeB stars of simulated BSP with $\eta$
= 0.5 and the distributions of the $\nu_{max}$ and $\Delta\nu$ as a
function of stellar age. The distributions of \textbf{the}
$\nu_{max}$ and $\Delta\nu$ are very similar to those of the SSP
with the same $\eta$. This is because that most binary stars evolved
into CHeB stage are wide binary stars. Binary interactions such as
mass transfer and mass accretion do not take effect in these wide
binary stars. However, the lower panels of Fig. \ref{fbsp} show that
the value of $\nu_{max}$ and $\Delta\nu$ of some stars with age $>$
2 Gyr is larger than 40 $\mu$Hz and 4 $\mu$Hz, respectively. This is
due to the fact that binary interactions lead to an obvious increase
or decrease in the mass of the stars when they evolve into the CHeB
stage. Consequently, the $\nu_{max}$ and $\Delta\nu$ of these stars
are larger than those of the wide binary stars. However, the
fraction of these stars is very small. Thus the effect of binary
interactions on the distributions of $\nu_{max}$ and $\Delta\nu$ is
not significant.

\subsection{SFR effect}
\begin{figure}
\begin{center}
\includegraphics[angle=-90, width=0.47\textwidth]{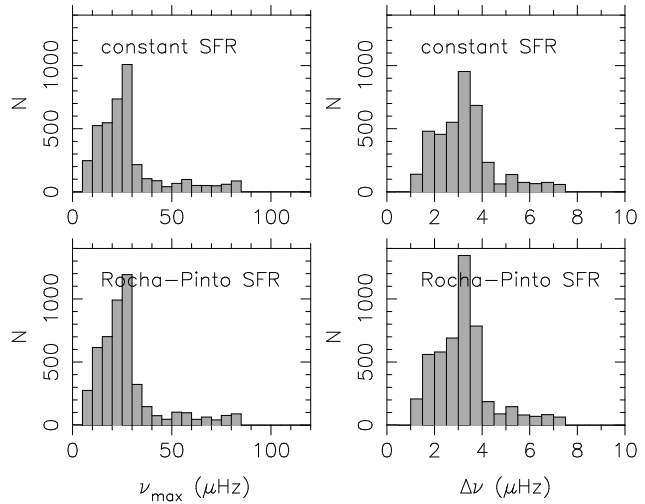}
 \caption{The histograms of the $\nu_{max}$ and $\Delta\nu$ of
 simulated CHeB stars for different star formation rate.
 Here, $\eta$ = 0.5. Upper panels show the
 results of SSP with constant SFR. While lower panels represent the
 results of SSP with the Galaxy SFR given by \citet{roch00b}.}
 \label{fsfr}
\end{center}
\end{figure}

The star formation rate of the Galaxy is not a constant.
\citet{roch00b} gave that the SFR of the Galaxy at 2-5 Gyr ago is
about 2 times larger than an average SFR over the past 15 Gyr and
the SFR of the Galaxy at 7-9 Gyr ago is around 1.5 times larger than
the average SFR. Fig. \ref{fsfr} shows the distributions of
$\nu_{max}$ and $\Delta\nu$ of CHeB stars of the simulated SSP with
a constant SFR and with the Galaxy SFR given by \citet{roch00b}.
Although the number of simulated CHeB stars is changed, the
distributions are similar. In fact, for $\eta$ = 0.5, Fig.
\ref{faged} shows that the value of \textbf{the} $\nu_{max}$ and
$\Delta\nu$ of the CHeB stars with age $>$ 2 Gyr is mainly located
in the range of 10-30 $\mu$Hz and 1-4 $\mu$Hz, and gathers around 25
$\mu$Hz and 3 $\mu$Hz, respectively. Thus increasing or decreasing
the SFR of stars at a certain age $>$ 2 Gyr can increase or decrease
the number of CHeB stars but \textbf{cannot} affect the peak
\textbf{locations of the} $\nu_{max}$ and $\Delta\nu$ of CHeB stars.
\textbf{Although} increasing the SFR of the young stars \textbf{in
the age range of} 0-2 Gyr \textbf{also} can increase the number of
stars with relatively high $\nu_{max}$ and $\Delta\nu$, \textbf{the
peak locations cannot also be affected} unless the SFR is enhanced
many times.

\subsection{Mixing-length parameter effect}

\begin{figure}
\begin{center}
\includegraphics[angle=-90, width=0.47\textwidth]{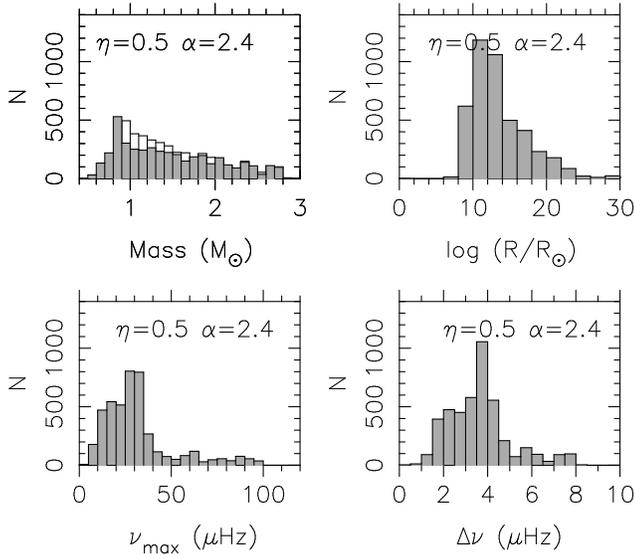}
 \caption{The histograms of the $\nu_{max}$, $\Delta\nu$, M, and R
 of the CHeB stars of simulated SSP with the mixing-length parameter
 $\alpha$ = 2.4.}
 \label{fpa24}
\end{center}
\end{figure}

\begin{figure}
\begin{center}
\includegraphics[angle=-90, width=0.4\textwidth]{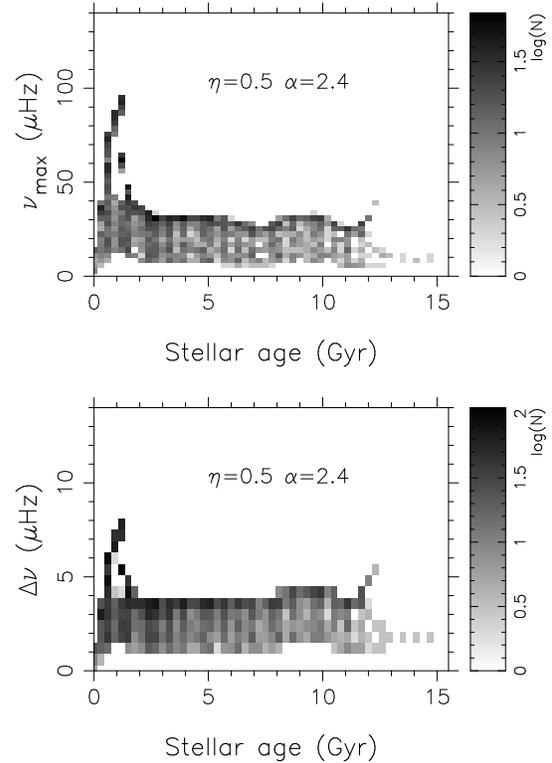}
 \caption{The distributions of the $\nu_{max}$ and $\Delta\nu$ of
 the CHeB stars of simulated SSP with $\alpha$ = 2.4 as a
 function of stellar age.}
 \label{fag24}
\end{center}
\end{figure}

The \textbf{distribution of the radius and of the $\Delta\nu$} of
simulated CHeB stars have a dominant peak, but the distribution of
\textbf{the} mass of the simulated CHeB stars with $\eta$ = 2.0 has
not a dominant peak, which implies that the non-uniform
distributions of $\nu_{max}$ and $\Delta\nu$ may be caused by the
radius of many CHeB stars concentrating in a narrow interval. In
fact, equations \ref{eq1} and \ref{eq2} indicate that $\nu_{max}$
and $\Delta\nu$ are more sensitive to the change in radius than that
in mass. Furthermore, a variation of the mixing-length parameter
$\alpha$ mainly changes the stellar radius, but has almost no
influence on the luminosity \citep{kipp90}. Used the Eggleton's
stellar evolution code \citep{egg71,egg72,egg73}, our calculations
show that the increase in $\alpha$ leads to a decrease in the radius
of stars evolved into horizontal branch, and thus leads to an
increase in \textbf{the} $\nu_{max}$ and $\Delta\nu$ of the stars.
In order to study the effect of the mixing-length parameter $\alpha$
on the distributions of $\nu_{max}$ and $\Delta\nu$, we modified the
Hurley's single evolution code to calculate the stellar evolution
with $\alpha$ = 2.4. Fig. \ref{fpa24} shows the histograms of
\textbf{the} mass, radius, $\nu_{max}$ and $\Delta\nu$ of the
simulated CHeB stars with $\alpha$ = 2.4. \textbf{An} increase in
$\alpha$ results in a movement of the peak location of the radius
distribution towards a lower value, \textbf{and} leads to a movement
towards a higher value for the peak locations of the $\nu_{max}$ and
$\Delta\nu$ distributions. Furthermore, Fig. \ref{fag24} represents
the distributions of the $\nu_{max}$ and $\Delta\nu$ as a function
of stellar age. Increasing the mixing-length parameter $\alpha$
leads to the upper boundary of the distributions moving up slightly.

\section{Discussions and Conclusions}

Several points should be kept in mind when comparing the theoretical
distributions of $\nu_{max}$ and $\Delta\nu$ with the observations.
Firstly, the scaling equation (\ref{eq1}) is obtained from the ideas
that $\nu_{max}$ is proportional to the acoustic cutoff frequency
$\nu_{ac}$ and $\nu_{ac} \propto gT^{-1/2}_{eff}$ \citep{brow91},
where $g$ is the stellar surface gravity and $T_{eff}$ is the
effective temperature. The $\nu_{max}$ given by equation (\ref{eq1})
is not necessarily accurate \citep{gill08, stel09a}. For example,
the ratio of $\nu_{ac}$ of stellar models to that calculated from
the global parameters of models is always below unity ( see the
Figure 8 of \cite{stel09a}). However, the scaling equation
(\ref{eq2}) is very accurate. The uncertainty of equation
(\ref{eq2}) is within a few per cent \citep{stel09a, stel09b}. In
addition, scaling equations (\ref{eq1}) and (\ref{eq2}) agree within
a few per cent with stellar model calculations for cool models
($T_{eff} \lesssim$ 6400 K) \citep{stel09a}. A few per cent
deviations in equations (\ref{eq1}) and (\ref{eq2}) \textbf{cannot}
obviously change the peak locations of $\nu_{max}$ and $\Delta\nu$
in our simulations. \textbf{Hence,} the deviations \textbf{cannot}
affect our basic results.

The giant stars observed by the CoRoT may contain the first giant
branch \textbf{(FGB)} stars. Our simulations show that the
distributions of $\nu_{max}$ and $\Delta\nu$ of the \textbf{FGB
stars} are uniform. \textbf{Therefore,} the $\nu_{max}$ and
$\Delta\nu$ of the \textbf{FGB stars} do not affect the peak
locations.

Comparing with the \citet{migl09}'s simulation, we used the same IMF
and age-metallicity relation. In our simulation, the SFR is a
constant over the past 15 Gyr. But it is a constant over the past 9
Gyr in the \citet{migl09}'s simulation. This difference could not
affect the simulated results because the oldest population are
impeded into the CHeB stage by mass loss. For example, for $\eta$ =
1.0, the simulated population do not contain the stars older than
about 10 Gyr. The frequencies of our simulated peak locations are
lower than those of \citet{migl09}. This may be caused by
mixing-length parameter.

For different $\eta$, the CHeB stars with age $<$ 2 Gyr lose only a
negligible amount of mass from their surface before they evolved
into CHeB stage. \textbf{Consequently,} the effect of the mass loss
on the evolution of these stars is insignificant. The distributions
of $\nu_{max}$ and $\Delta\nu$ of these stars are almost not
affected by Reimers mass loss. \textbf{However,} for old population,
the bigger the $\eta$, the more the mass loss. For a big $\eta$,
stars \textbf{can} lose an appreciable, but from star to star
different, amount of mass from their surface before the helium
flash. Then the stars move to the left of the HR diagram with
slightly decreasing luminosity after the flash, and the mean density
of the stars is enhanced. The increase in the mean density leads to
\textbf{an increase in the $\nu_{max}$ and $\Delta\nu$ of these
stars}. However, the change in the mean density is different from
star to star. The mass of the stars evolved into CHeB at the same
age is approximate, but the extent of central helium burning is
different. The stars around the ZAHB have a smaller radius than
\textbf{those} approaching the AGB. The fractional change caused by
mass loss in the radius of the stars around the ZAHB is larger, and
the change in the mean density of these stars is also larger than
\textbf{that of those} approaching the AGB. \textbf{The} stars which
are close to the ZAHB are located around the upper boundary of the
distributions of $\nu_{max}$ and $\Delta\nu$ with age, while
\textbf{those} approaching the AGB are located around the lower
boundary. \textbf{Therefore,} the changes of the upper and lower
boundary shown in Fig. \ref{faged} are different. However, for the
middle age population, the stars lose only a little amount of mass
from their surface before the helium flash, which \textbf{hardly}
affect the radius and luminosity of the stars after the flash.
Because the mass decrease slightly but the radius is almost not
changed, the mass loss results in that the $\nu_{max}$ and
$\Delta\nu$ of these stars decrease slightly.

The Reimers mass loss mainly affects the old population. The mass
loss leads to an increase in the $\nu_{max}$ and $\Delta\nu$ of old
population. \textbf{However,} a high mass-loss rate can impede the
low-mass low-metal stars evolving into CHeB stage. If the Reimers
mass loss is very efficient during the red-giant branch, the
$\nu_{max}$ and $\Delta\nu$ of CHeB stars would not be observed in
old clusters. The asteroseismical observation on old clusters may
provide a help to constrain the mass-loss rate.

For $\eta$ = 0.5, the value of $\nu_{max}$ and $\Delta\nu$ of
population with age $>$ 2 Gyr is almost located in the range of
10-30 $\mu$Hz and 1-4 $\mu$Hz, and gathers about 25 $\mu$Hz and 3
$\mu$Hz, respectively. \textbf{Therefore} increasing or decreasing
the number of stars at a certain age $>$ 2 Gyr \textbf{cannot}
affect the peak locations of the $\nu_{max}$ and $\Delta\nu$. So the
peak locations are not sensitive to the SFR and \textbf{not
sensitive to} whether population contains old stars. For $\eta$ =
1.0, increasing the SFR of stars with age between 7-9.5 Gyr can
increase the stars with $\nu_{max} >$ 25 $\mu$Hz and $\Delta\nu
>$ 3 $\mu$Hz. But even enhancing the SFR to several times, the peak
locations are not affected.

In BSP, some CHeB stars can lose a little mass, but some stars can
accrete a little mass by the weak binary interactions. The effect of
\textbf{a slight mass change} before the helium flash on the
luminosity and radius of CHeB stars after the flash is negligible.
\textbf{Therefore} the $\nu_{max}$ and $\Delta\nu$ of the CHeB stars
which lost a little bit of mass decrease; whereas those of the CHeB
stars which accreted a little mass increase. \textbf{Hence} the
distributions of $\nu_{max}$ and $\Delta\nu$ of CHeB stars
\textbf{cannot} be affected. However, for the strong binary
interactions, on the one hand, some stars which have lost an
appreciable amount of mass from their surface would move to the left
of the H-R diagram with slightly decreasing luminosity, at the same
time, the mean density of these stars would increase.
\textbf{Consequently} the $\nu_{max}$ and $\Delta\nu$ of these stars
increase. \textbf{On} the other hand, some stars with age $>$ 2 Gyr
which have accreted a considerable amount of mass would become like
the stars with age $<$ 2 Gyr, \textbf{hence} their $\nu_{max}$ and
$\Delta\nu$ can increase obviously too. However, the fraction of
these interactive binary stars appearing in our simulated CHeB stars
is very small. \textbf{Therefore}, although the binary interactions
such as mass transfer and mass accretion can affect the $\nu_{max}$
and $\Delta\nu$ of CHeB stars undergoing a mass accretion or mass
loss, the effect of binary interactions on the distributions of
\textbf{the} $\nu_{max}$ and $\Delta\nu$ of CHeB stars is not
significant.

An increase in the mixing-length parameter $\alpha$ mainly leads to
a decrease in the radius of all CHeB stars after the helium flash.
Thus the $\nu_{max}$ and $\Delta\nu$ of CHeB stars increase with
$\alpha$. Consequently, the peak locations of the $\nu_{max}$ and
$\Delta\nu$ can be affected by the $\alpha$. For the middle-age
population \textbf{at the age of about 2 to 5 Gyr}, the peak
locations are more sensitive to the $\alpha$ than mass-loss rate,
SFR and metal abundance. Thus the asteroseismical observation on the
middle age clusters may provide a help to constrain the
mixing-length parameter.

The mass of simulated CHeB stars is mainly located in 1-2
$M_{\odot}$. For the CHeB stars with the mixing-length parameter
$\alpha$ = 2.0, \textbf{the most of them} have a radius \textbf{in
the range of} 11-14 $R_{\odot}$. \textbf{Therefore,} even the mass
distribution is uniform, the stars have an approximate mean density.
\textbf{Hence} they have an approximate $\nu_{max}$ and $\Delta\nu$.
Consequently, there is a dominant peak in the distributions of the
$\nu_{max}$ and $\Delta\nu$.

A high Reimers mass loss can lead to an increase in the $\nu_{max}$
and $\Delta\nu$ of old population, and impedes the low-mass
low-metal stars evolving into CHeB stage. \textbf{On  the contrary,
it results in a very small decrease in those of middle-age
population}. However, the mass loss \textbf{scarcely} affect the
$\nu_{max}$ and $\Delta\nu$ of young population. The effect of the
Reimers mass loss on the peak locations of $\nu_{max}$ and
$\Delta\nu$ of the CHeB stars is not significant unless the
mass-loss rate is very high. The effect of star formation rate and
binary interactions on the peak locations of $\nu_{max}$ and
$\Delta\nu$ of the CHeB stars is also not significant. The dominant
peak of $\nu_{max}$ and $\Delta\nu$ is due to the fact that most of
CHeB stars have an approximate radius. The radius can be affected by
the mixing-length parameter. \textbf{The} peak location also can,
\textbf{thus}, be affected by the mixing-length parameter.

\section*{Acknowledgments}
We thank \textbf{Shaolan Bi for her help,} the anonymous referee for
his/her helpful comments. This work was supported by the Ministry of
Science and Technology of the People¡¯s republic of China through
grant 2007CB815406, the NSFC though grants 10773003, 10933002,
10963001, \textbf{project of the fundamental and frontier research
of henan province under grant no. 102300410223}, and the
high-performance grid computing platform of Henan Polytechnic
University.

\end{document}